\documentclass[letterpaper,twocolumn,10pt]{article}
\usepackage{usenix2024_SOUPS}

\usepackage{tikz}
\usepackage{amsmath}

\usepackage{filecontents}

\usepackage{dirtytalk}

\begin{document}

\date{}

\title{\Large \bf Privacy Risks of General-Purpose AI Systems:\\
  A Foundation for Investigating Practitioner Perspectives}

\def\plainauthor{Stephen Meisenbacher, Alexandra Klymenko, Patrick Gage Kelley, Sai Teja Peddinti, Kurt Thomas, Florian Matthes}

\author{
{\rm Stephen Meisenbacher}\\
Technical University of Munich
\and
{\rm Alexandra Klymenko}\\
Technical University of Munich
\and
{\rm Patrick Gage Kelley}\\
Google
\and
{\rm Sai Teja Peddinti}\\
Google
\and
{\rm Kurt Thomas}\\
Google
\and
{\rm Florian Matthes}\\
Technical University of Munich
} 


\maketitle
\thecopyright

\begin{abstract}
The rise of powerful AI models, more formally \textit{General-Purpose AI Systems} (GPAIS), has led to impressive leaps in performance across a wide range of tasks. At the same time, researchers and practitioners alike have raised a number of privacy concerns, resulting in a wealth of literature covering various privacy risks and vulnerabilities of AI models. Works surveying such risks provide differing focuses, leading to disparate sets of privacy risks with no clear unifying taxonomy. We conduct a systematic review of these survey papers to provide a concise and usable overview of privacy risks in GPAIS, as well as proposed mitigation strategies. The developed privacy framework strives to unify the identified privacy risks and mitigations at a technical level that is accessible to non-experts. This serves as the basis for a practitioner-focused interview study to assess technical stakeholder perceptions of privacy risks and mitigations in GPAIS. 
\end{abstract}

\section{Introduction}
\label{sec:introduction}

Recent advances in Artificial Intelligence have fueled the rapid proliferation of AI systems. Such systems require careful data curation and the employment of state-of-the-art training procedures. With the promise of AI, specifically General-Purpose AI Systems (GPAIS), follows a number of risks to privacy. The sheer amount of data involved, as well as errors arising in development and deployment, open the door for privacy vulnerabilities that can transpire into privacy breaches. Such risks may make developers reluctant to deploy AI technologies amidst concerns about safe and responsible AI usage.

There exist several recent works in the academic literature that survey the landscape of AI privacy risks, particularly in the field of LLMs. The sets of risks presented in these works are quite disjoint, with each work presenting a different perspective. Furthermore, existing frameworks focus on providing an \textit{academic} understanding of the AI privacy risk landscape, leading to limited practical usability without a clear mapping of mitigations to risks, as well as lacking insight into practitioner readiness to implement such mitigations.

In this work, we present a systematic overview of privacy risks associated with the development, deployment, and application of GPAIS, as well as the existing mitigation strategies to combat these risks. We plan to bundle these insights into a \textit{risk catalog} for practitioners, a guide for deciding which risks and mitigations are most relevant to one's personal role.  

We hope that our work will raise awareness of the potential privacy risks of AI systems
by addressing a gap in the literature that existing proposed taxonomies missed, namely a practical perspective. Thus, we strive to create a risk catalog for dissemination so that more developers may be aware of which privacy risks affect their own work, and what can be done about it. In particular, we see it as very important that technical professionals are included in the design and creation of our catalog, as these are the people working day-to-day on the development of these powerful, yet risky, technologies. As such, the implications for society rest largely in the hands of the technical people working on GPAIS, and through an interview study, we hope to capture this crucial perspective.

As our findings are in an early stage of refinement, we seek to gauge the practical efficacy of our catalog in development, with a focus on aligning the contents with the needs of practitioners who wish to gain a better understanding of the risks posed by GPAIS in a technically-founded yet understandable way, as well as of what they can do to mitigate these risks.


\section{Methodology}
\label{sec:methodology}
Our work is guided by three research questions, namely:

\begin{itemize}
    \itemsep -0.1em
    \item[RQ1] What are the privacy risks of General-Purpose AI systems and how can they be systematized in a way that is practically usable and helpful to technical stakeholders?
    \item[RQ2] What solutions exist to mitigate the identified privacy risks in GPAIS?
    \item[RQ3] What are technical stakeholder perceptions of privacy risks in GPAIS and their possible mitigations? 
\end{itemize}

\subsection{Ongoing Work}

To answer RQ1 and RQ2, we conduct a systematic literature review on the topic of privacy risks in AI. Following the methodology proposed by Kitchenham \cite{kitchenham2015evidence}, we define the following search string (for titles only):
\begin{center}
    \scriptsize
    \textit{("privacy" OR "private") AND ("risks" OR "risk" OR "harms" OR "harm" OR "threats" OR "threat" OR "concerns" OR "concern" OR "dangers" OR "danger" OR "protect*" OR "mitigat*") AND ("AI" OR "Artificial Intelligence" OR "GPAIS" OR "General Purpose AI" OR "General Purpose Artificial Intelligence" OR "Language Model" OR "LLMs" OR "Generative AI" OR "Diffusion Model" OR "Multimodal Model")}
\end{center}


We utilize this search string in Google Scholar. This choice was limited due to the comprehensiveness of Google Scholar, as it encompasses many other prominent outlets such as ACM Digital Library, IEEExplore, and ACL Anthology. 

Using the above search string yields 89 results. This was initially filtered to 87 by only considering papers from 2015 onwards. Next, title and abstract screening was performed to remove the following categories of results: (1) not accessible via public or institutional login, or (2) specific case study or implementation proposal. The second point was included in order to emphasize papers of the \textit{survey} nature.

Initial filtering resulted in a core set of 5 papers. From this, we performed forward/backward search to find other relevant works, as well as added in gray literature sources already in our possession \cite{garousi2019guidelines}. This led to a final set of \textbf{12} survey papers ~\cite{council2023ghost, rigaki2023survey, curzon2021privacy, shahriar2023survey, vassilev2024adversarial, yan2024protecting, bsi2024ai, biggio2018wild, smith2023identifying, golda2024privacy, rahman2023security, mcgraw2024architectural}. These sources were analyzed by our research team, resulting in an initial set of privacy risks and mitigations, presented in Section \ref{sec:findings}.
These serve as the basis for answering RQ3, which primarily involves collecting expert feedback.

\subsection{Next Steps: Expert Feedback}
Our study revolves around listening to practitioner voices in the creation of a GPAIS privacy risk taxonomy that includes mitigations. As such, we hope to expand beyond a literature search by augmenting our initial findings with expert feedback. We aim to do this in two formats: (1) live expert feedback, and (2) semi-structured interviews with technical professionals.

\paragraph{Workshop Feedback.}
We hope to use the SUPA workshop format for the presentation of our initial results and to receive open feedback.
Our hope is to lead a discussion surrounding our initial findings for RQ1 and RQ2, and to discuss open questions, presented in Section \ref{sec:open_questions}. The particular venue of SUPA, focusing on user-centered privacy, presents an excellent opportunity to do so. From this, we aim to identify remaining gaps in our findings and to collect further insights for the analysis stage of our interview study.

\paragraph{Semi-structured Interviews.}
In the summer of 2024, we plan to conduct a series of semi-structured interviews with technical professionals working in the field of AI, where the primary goal will be to gauge their perception of the identified set of privacy risks and mitigations.
Interview candidates will be acquired through personal contacts, online platforms such as LinkedIn, and references provided by interviewees in the study itself. Additionally, contacts acquired through venues like SOUPS and SUPA will be very valuable for the study. 

We have already created an interview guide for the study, which explores which privacy risks of GPAIS the interviewee is aware of, but more importantly, how these risks are \textit{perceived}. A crucial element of the interviews will be a ranking exercise in which the interviewee ranks the importance of all presented risks. In addition, we aim to measure the perceived readiness of the developers to \textit{mitigate} these risks, and in parallel, to determine what the interviewee believes to be missing in privacy risk mitigation strategies.


\section{Preliminary Privacy Framework}
\label{sec:findings}

\subsection{Privacy Risks in GPAIS}
\label{sec:risks}

Below, we introduce the identified privacy risks, grouped into six categories.
In the final catalog, for each risk, we plan to include the following information: \textit{risk name} and \textit{description}, \textit{risk type }(inherent/implementation/interface, i.e., to where the risk can be attributed), \textit{risk source} (data-/ model-/system-level), \textit{lifecycle stage} where the risk originates (development/deployment/application), specific \textit{examples} for the risk, and mapping to applicable \textit{mitigations}. 

\paragraph{Data Management.}
 Data management risks involve those associated with the collection, storage, and sharing of raw training data for developing AI systems. Risk sources include lack of consent, improper handling or storage, improper sharing with third parties, and errors in processing. \cite{shahriar2023survey}

\paragraph{Data Memorization and Leakage.}
Larger models have the tendency to “memorize” training data, particularly the data that is more unique or rare-occurring, and this issue is made more dangerous with the ability to “prompt” models. Data leakage occurs when models unintentionally expose their memorized data, and in the case of prompting, malicious users may be able to reconstruct or extract data. This may even occur when text is released in embedding form. \cite{bsi2024ai,rahman2023security, shahriar2023survey, smith2023identifying, vassilev2024adversarial}


\paragraph{Unintended Downstream Usage of Sensitive Prompts and User Metadata.}
User inputs and queries may be used for originally unintended purposes, e.g. building up user profiles based on contextual information, or sensitive company data being captured and used downstream for model fine-tuning.

\begin{itemize}
    \itemsep -0.1em
    \item \textit{Profiling/Contextualization}: As interaction with many GPAIS takes the form of \textit{prompting} the models, users often contextualize their queries to optimize the outputs given by a model. In providing information that gives context about a user, there exist the risks of these AI systems building a \say{profile} of the user, including sensitive or otherwise private information.
    \cite{bsi2024ai,yan2024protecting}
    \item \textit{Sensitive or Protected Queries}: A known risk of interacting with GPAIS is the input of private, protected, or confidential data, such as enterprise information. Especially in situations where a system stores input prompts for future fine-tuning, one may be wary to share their data with model endpoints in order to avoid data leaks. \cite{yan2024protecting}
\end{itemize}

\paragraph{Adversarial Inference and Inversion.}
A class of attacks referred to as \textit{inference} attacks includes an adversary trying to distinguish whether a particular instance (member) is in the training data, attempting to recover attributes of the training data, making broader inferences about the underlying data, or reverse-engineering model-specific parameters.

\begin{itemize}
    \itemsep -0.1em
    \item \textit{Membership Inference}: This attack aims to infer whether a particular member (user or training instance) was present in the original training set. Such gained information is dangerous if knowledge of the data domain or purpose is also known, whereby attackers are able to infer further information about an individual. \cite{rahman2023security,rigaki2023survey, shahriar2023survey, smith2023identifying,vassilev2024adversarial, yan2024protecting}
    \item \textit{Attribute Inference}: This adversarial attack goes one step further, trying to learn about the attributes, or features, of a training instance. This can materialize in different ways, such as learning structured features of a training instance, or in the case of natural language, learning about the makeup of input text. \cite{jayaraman2022attribute, rigaki2023survey, shahriar2023survey, yan2024protecting}
    \item \textit{Property Inference}: Malicious attackers may also wish to learn about the \textit{properties} of the training data, drawing conclusions about the distribution or demographics of the data. This is harmful in leaking information about the general characteristics of a dataset, such as the percentage of instances containing a sensitive attribute. \cite{rigaki2023survey, vassilev2024adversarial}
    \item \textit{Model Inversion}: When models are exposed to the public, this requires opening query access via some endpoint. Given unlimited access to a model, malicious users may try to infer information about a model’s parameters or architecture, purely from observing the answers to queries posed to the model. In this, the danger arises when these users are able to \say{invert} or \say{steal} a model. \cite{bsi2024ai,rahman2023security, shahriar2023survey, smith2023identifying,vassilev2024adversarial, yan2024protecting}

\end{itemize}

\paragraph{Misuse via Harmful Applications.}
The capabilities of GPAIS open the door for misuse in downstream applications, where malicious users leverage AI for nefarious purposes (e.g., deepfakes, malicious code or phishing email generation, deceiving AI-generated text detectors, etc.). \cite{council2023ghost,golda2024privacy,mcgraw2024architectural}

\paragraph{Other.} Beyond the set of technically related risks, we formulate two further risk categories that exist at an ecosystem level. In particular, we emphasize two higher categories, namely \textit{legal and regulatory} and \textit{ethical and societal}. 

\begin{itemize}
    \itemsep -0.1em
    \item \textit{Legal and Regulatory}: The mandate handed down by modern privacy laws and regulations is intended to safeguard the privacy of individuals with regard to the collection and use of personal data. This includes providing clear privacy policies and terms of service. When AI systems do not properly adhere to these guidelines, privacy is at risk. This also holds financial implications for AI developers (i.e., the companies behind them). \cite{shahriar2023survey}
    \item \textit{Ethical and Societal}: The ethical and societal implications of AI systems with respect to privacy must also be considered. As such systems hold considerable power to affect the daily lives of people, the impact of privacy breaches must be considered paramount in the design, development, and deployment of AI systems. \cite{shahriar2023survey, council2023ghost}
\end{itemize}

\subsection{Mitigations}
\label{sec:mitigations}

The mitigations are organized into five categories and are briefly presented below. In the catalog, in addition to its \textit{description} and \textit{mapping to risks}, we plan to include \textit{tangible available examples} of the mitigation, its \textit{benefits and limitations}, as well as possible \textit{combinations} with other mitigations.

\paragraph{Organizational}

\begin{itemize}
    \itemsep -0.1em
    \item \textit{Management of (Training) Data}: Mitigating privacy risks begins with the responsible handling of training data, for example in trusted and secured data warehouses, as well as proper data governance structures. \cite{bsi2024ai, golda2024privacy}
    \item \textit{Alignment}: An important organizational measure involves the \textit{aligning} of GPAIS to human values, a step that can help to ensure trustworthiness and reduce the risk of adversaries compromising models. \cite{bsi2024ai}
    \item \textit{Trusted Execution Environments/Confidential Computing}: Secured hardware environments where data processed within cannot be read or tampered with by outside parties or code. \cite{curzon2021privacy, yan2024protecting}
\end{itemize}

\paragraph{Data Preprocessing}
\begin{itemize}
    \itemsep -0.1em
    \item \textit{Input Sanitization/Data Cleaning}: When preparing data for model training, a wise preprocessing step includes data cleaning, in which explicit sensitive information is removed, such as phone numbers. \cite{curzon2021privacy,bsi2024ai,rahman2023security, smith2023identifying ,yan2024protecting}
    \item \textit{Anonymization/Pseudonymization}: Similarly, explicit personally identifiable information (PII) can be removed via anonymization or pseudonymization techniques. \cite{bsi2024ai} 
    \item \textit{Outlier Detection}: Outlier detection methods aim to identify and detect outliers in the training data, as these data points might be more readily memorized. \cite{mcgraw2024architectural,shahriar2023survey,vassilev2024adversarial}
\end{itemize}

\paragraph{Data Augmentation}
\begin{itemize}
    \itemsep -0.1em
    \item \textit{Differential Privacy (DP) (Randomized Response)}: DP is a mathematically grounded notion of privacy which usually involves adding random noise to query outputs to inject plausible deniability into computations on potentially private data (attributes). \cite{curzon2021privacy,bsi2024ai,golda2024privacy,smith2023identifying,vassilev2024adversarial, yan2024protecting}  
    \item \textit{Group Privacy}: The notions of k-anonymity, l-diversity, and t-closeness concern themselves with \textit{group privacy}, where the goal is to provide a certain level of indistinguishability between members of a dataset. \cite{council2023ghost,curzon2021privacy,vassilev2024adversarial}
    \item \textit{Data Transformation}: Storing data in \say{transformed} forms, such as in embedding format or in reduced dimensions, may help to obfuscate raw sensitive data. \cite{curzon2021privacy,shahriar2023survey}
    \item \textit{Privacy-Preserving Data Aggregation/Homomorphic Encryption}: Advanced privacy-preserving techniques based on encryption can be leveraged such that computation on raw user data is not necessary. \cite{golda2024privacy,yan2024protecting}
    \item \textit{Functional Secret Sharing (Secure Multiparty Computation)}: In some privacy-preserving mechanisms, user data is \say{split} among a pool of users, such that no single data point must be shared in full. \cite{yan2024protecting}
    \item \textit{Synthetic Data}: Rather than use the original training data, some training strategies may opt to use synthetically generated data, which ideally shares a similar distribution of the original data. \cite{curzon2021privacy}
\end{itemize}

\paragraph{Private Training}
\begin{itemize}
    \itemsep -0.1em
    \item \textit{Adversarial Training}: Adding augmented samples to training data can help to improve model robustness, thereby reducing the risk of privacy breaches. \cite{curzon2021privacy,bsi2024ai,rahman2023security}
    \item \textit{Model Diversification}: Some training procedures may opt to train a variety of models (or parameters), not only to improve robustness in decision-making, but also to mitigate any vulnerabilities of one single model. \cite{rahman2023security}
    \item \textit{Federated Learning}: The training of models is performed locally in a \textit{distributed} fashion, where only model updates are shared with a central aggregator. \cite{golda2024privacy,rahman2023security, yan2024protecting}
\end{itemize}

\paragraph{Post Hoc Privatization}
\begin{itemize}
    \itemsep -0.1em
    \item \textit{Model Explainability}: Focusing on the development of \textit{explainable} models can detect model vulnerabilities, as well as verify the integrity of a model. \cite{bsi2024ai,golda2024privacy,rahman2023security, shahriar2023survey}
    \item \textit{Interface-side Mitigations}: Safeguarding the interface between users and models includes detecting suspicious queries, limiting the number of queries, and building guardrails. Such measures serve to protect legitimate users and to reduce adversarial advantage. \cite{bsi2024ai,smith2023identifying,vassilev2024adversarial}
    \item \textit{Machine Unlearning}: This novel technique aims to provide developers with the ability to remove a single user's contributions to a training data set upon request, without the need for complete retraining. \cite{bsi2024ai,smith2023identifying ,yan2024protecting} 
    \item \textit{Private Inference}: Augmenting model inference (computation on unseen data) with privacy-preserving solutions such as Differential Privacy or Homomorphic Encryption adds an extra layer of privacy protection for users inputting potentially sensitive data. \cite{bsi2024ai}
\end{itemize}

\section{Discussion Points and Open Questions}
\label{sec:open_questions}
Here, we introduce a set of open questions, for which we aim to gain insights by engaging in discussions at the workshop.

\begin{enumerate}
    \itemsep -0.1em
    \item \textit{Are the presented lists of privacy risks and their possible mitigations well-categorized?} \\
   The question above poses whether the presented categorization of risks and mitigations is the most sensible, and if not, what improvements can be made.
    \item \textit{Are the presented lists of privacy risks and their possible mitigations comprehensive?} \\
We wish to discern whether the set of risks and mitigations presented under each of the categories is comprehensive, or if anything missing should be included.
    
    \item \textit{Is the proposed catalog structure usable? Beyond this, what extra dimensions would make our framework practically relevant and helpful to technical professionals?} \\
To make our privacy risk catalog as practically usable as possible, we would like to get feedback regarding its planned structure, presented in Sections \ref{sec:risks} and \ref{sec:mitigations}.
    
    \item \textit{In designing the catalog, how do we ensure its adaptability as new risks and mitigations arise?} \\
    We plan for the catalog to be a living resource, i.e. updated on a regular basis. As such, we look for feedback on best practices for managing this dynamic nature.


\end{enumerate}

\section{Conclusion}
We present our motivation, methodology, and working plan for constructing a GPAIS privacy risk and mitigation catalog, forming the foundation for investigating technical practitioners' perceptions of AI risks and their possible mitigations. With this, our goal is to discuss our work in progress at the SUPA workshop and to collect feedback regarding any points we may have missed. Guided by our initial results and open questions, we are confident that our ongoing work may spark interesting discussions and fruitful points for consideration.

\label{sec:conclusion }


\newpage
\bibliographystyle{plain}
\bibliography{usenix2024_SOUPS}


\end{document}